\documentclass[aps,prd,floatfix,superscriptaddress,notitlepage,onecolumn,showpacs]{revtex4}

\usepackage{epsfig}
\usepackage{amsmath}
\usepackage{graphicx,psfrag}
\usepackage{dcolumn}
\usepackage{bm}
\usepackage{slashed}
\usepackage{xcolor}
\usepackage{maybemath}
\usepackage[makeroom]{cancel}
\usepackage[normalem]{ulem}

\newcommand{\hf} {\frac{1}{2}}

\newcommand\fig[1]     {Fig.\,{\ref{#1}}}

\def\eq#1{(\ref{#1})}
\def\s0#1#2{\mbox{\small{$ \frac{#1}{#2} $}}}
\def\0#1#2{\frac{#1}{#2}}
\def\Eq#1{Eq.~(\ref{#1})}

\def\ord#1{{\cal O}(#1)}

\def\mr#1{{\mathrm{#1}}}

\usepackage{xcolor}

\definecolor{garrosgreen}{rgb}{0.1, 0.4, 0.1}
\definecolor{dartmouthgreen}{rgb}{0.05, 0.5, 0.06}
\definecolor{cambridgeblue}{rgb}{0.7, 0.2, 0.1}
\definecolor{cambridgeblue}{rgb}{0.1, 0.3, 1.0}
\definecolor{oxfordblue}{rgb}{0.05, 0.2, 0.7}
\definecolor{OliveGreen}{rgb}{0,0.6,0}
\definecolor{mig}{rgb}{0.0,0.7,0.0}

\newcommand{\rminf}{{i}}

\begin{document}

\newcommand{\addrmst}{Department of Physics, Missouri University of 
Science and Technology, Rolla, Missouri 65409, USA}

\newcommand{\addrmtade}{
MTA--DE Particle Physics Research Group, P.O.~Box 51, 
H--4001 Debrecen, Hungary}

\newcommand{\addreth}{Institut f\"ur Theoretische Physik, ETH Z\"urich, Wolfgang-Pauli-Str. 27 Zurich, Switzerland}

\newcommand{\addrdeb}{University of Debrecen, 
P.O.Box 105, H-4010 Debrecen, Hungary}

\newcommand{\addratomki}{MTA Atomki, P.O.~Box 51, H--4001 Debrecen, Hungary} 

\newcommand{\addunits}{Department of Physics, University of Trieste, Strada Costiera 11, I-34151 Trieste, Italy}

\newcommand{\addrcnr}{CNR-IOM DEMOCRITOS Simulation Center, 
Via Bonomea 265, I-34136 Trieste, Italy}

\newcommand{\addrsissa}{SISSA and INFN, Sezione di Trieste, 
Via Bonomea 265, I-34136 Trieste, Italy}

\title{Vacuum Energy and Renormalization of the Field-Independent Term}

\author{I. G. M\'ari\'an}

\affiliation{\addrmtade}
\affiliation{\addrdeb}

\author{U. D. Jentschura}

\affiliation{\addrmst}
\affiliation{\addrmtade}
\affiliation{\addratomki}

\author{N. Defenu}

\affiliation{\addreth}

\author{A. Trombettoni}

\affiliation{\addunits}
\affiliation{\addrcnr}
\affiliation{\addrsissa}

\author{I. N\'andori}

\affiliation{\addrmtade}
\affiliation{\addratomki}
\affiliation{\addrdeb}

\begin{abstract} 
Due to its construction, the nonperturbative renormalization
group (RG) evolution of the constant, field-independent term (which is constant
with respect to field variations but depends on the RG scale $k$) requires
special care within the Functional Renormalization Group (FRG) approach.
In several instances, the constant term of the potential has no
physical meaning. However, there are special cases where it receives important
applications. In low dimensions ($d=1$), in a quantum mechanical model, this
term is associated with the ground-state energy of the anharmonic oscillator.
In higher dimensions ($d=4$), it is identical to the $\Lambda$ term of the
Einstein equations and it plays a role in cosmic inflation. Thus, in
statistical field theory, in flat space, the constant term could be associated
with the free energy, while in curved space, it could be naturally associated
with the cosmological constant. It is known that one has to use a subtraction
method for the quantum anharmonic oscillator in $d=1$ to remove the $k^2$ 
term that appears in the RG flow in its high-energy (UV) limit in order to recover the
correct results for the ground-state energy. The subtraction is needed because
the Gaussian fixed point is missing in the RG flow once the constant term is
included.  However, if the Gaussian fixed point is there, no further
subtraction is required.  Here, we propose a subtraction method for $k^4$ and
$k^2$ terms of the UV scaling of the RG equations for $d=4$ dimensions if the
Gaussian fixed point is missing in the RG flow with the constant term. Finally,
comments on the application of our results to cosmological models are provided.
\end{abstract}


\maketitle

\tableofcontents

\section{Introduction}
\label{sec1}

The quantum field theory of gravity, i.e., Quantum Einstein Gravity (QEG) is
perturbatively non-renormalizable. It requires infinitely many unknown
parameters to be set by experiment.  A possible solution to perform
renormalization is the use of a nonperturbative treatment. Indeed,
nonperturbative renormalizability, which is also referred to as asymptotic
safety, provides us with a nontrivial high energy, i.e., ultraviolet (UV) fixed
point of the renormalization group (RG) flow. The RG flow leads to a finite
number of UV-attractive couplings; so, it is sufficient to perform only
a finite number of measurements. In other words, it controls the
UV behavior of the dimensionless couplings. They do not need to be small or
tend to zero in the UV limit but tend to finite values at the nontrivial UV
fixed point. It was shown in \cite{reuter} that for the simplest truncation of
QEG which is the Einstein--Hilbert action, such a nontrivial fixed point is
indeed present. Up to the present, many different works have already confirmed
that the asymptotic safety scenario is possible.  For a recent
review, we refer to~\cite{reuter_saueressig}. For applications to cosmology,
one consults Ref.~\cite{bonanno_saueressig,alessia}, and for applications to
black-hole physics and instructive explicit functional renormalization-group
computations, we refer to \cite{cosmo_qeg,quadratic_gravity,qeg1,qeg2,qeg3,%
qeg4,qeg5,qeg6,qeg7,qeg8,qegIR,matter_matters}.

Asymptotic safety is the concept that a theory is UV completed by
an interacting theory which, in the RG formalism, is described by a nontrivial
(non-Gaussian) UV fixed point and includes the assumption that 
the RG formalism can be used to solve the so-called
cosmological constant problem, too. One can use different methods to test this
assumption. The non-perturbative RG is a possible method, but there are also
others, including lattice and Monte Carlo methods.  In general, the
nonperturbative RG method \cite{WeHo1973,Po1984,eea_rg_1,eea_rg_2,AlPo2001} has
been used successfully in many areas of physics from statistical mechanics to
high energy physics
\cite{BeTeWe2002,Po2003,Gi2006,Vacca_Zambelli,qanharmonic,Pa2007,De2007,Ro2012,Na2014}
and in cosmology
\cite{rg_cosmo_1,rg_cosmo_2,rg_cosmo_6,rg_cosmo_7,rg_cosmo_10,rg_cosmo_12,
rg_cosmo_17,rg_cosmo_18} with particular attention to the cosmological constant
problem
\cite{rg_cosmo_constant_1,rg_cosmo_constant_2,rg_cosmo_constant_3,rg_cosmo_constant_6,
rg_cosmo_constant_10,rg_cosmo_constant_13,cosmo_constant_1,cosmo_constant_2,
cosmo_constant_3,cosmo_constant_5,cosmo_constant_6,cosmo_constant_7,
cosmo_constant_9,cosmo_constant_10,cosmo_constant_11,cosmo_constant_17,
cosmo_constant_19,Ru2020,MoSo2020}.  It is obvious that the vacuum energy
density induced by the quantum fluctuations would add to the cosmological
constant (see Ref.~\cite{Zeldovich_1,Zeldovich_2}). However, the calculated
vacuum energy density is larger by many orders of magnitude than the observed
cosmological constant; this is the essence of the cosmological constant problem
\cite{CC_1,CC_2}. Also, it is well known that the bare values of a physical
theory are connected to the low-energy phenomenology by RG transformations
\cite{Wilson_1,Wilson_2,Wilson_3,Zinn-Justin_1,Zinn-Justin_2,Zinn-Justin_3}.
So, if one could establish a connection of the bare value of the cosmological
constant to its low-energy equivalent by an RG method, then one could possibly
solve the problem: The bare value of the cosmological constant carries no
physical meaning and is treated as a running parameter. Indeed, in principle,
RG running can be used to connect its UV value to the observed low energy,
i.e., infrared (IR) one. Thus, in the functional renormalization group (FRG)
treatment of gravity, one sees that the cosmological constant corresponds to a
relevant direction. It means, there is no cosmological constant problem in the
nonperturbative treatment of gravity; it is a free parameter to be fixed by
observations.  However, let us note that one has to distinguish
between the RG scale $k$ and the observational scale, which we denote as $p$.
One should compute the physical limit $k\to 0$ by the RG method first and then
compute observables using the full quantum effective action.  Unfortunately,
this is not yet possible in reasonable approximations, so, one has a
qualitative understanding of the implications of asymptotically safe gravity,
assuming that the artificial flow with $k$ is a good approximation to the
physical flow with respect to the physical momentum $p$, which one can extract
from the effective action in the limit $k \to 0$. The flow generated by the FRG
equation can be analysed in regard to the question of whether it allows
trajectories that are compatible with the observational constraints on the
parameters of the action, since it is related to the $k$-flow and not to the
$p$-flow. In principle, the effective action, in the limit $k \to 0$, has
infinitely many parameters, but only $N$ of these parameters (those
associated with the $N$ relevant directions of the non-Gaussian UV fixed
point) must be fixed using $N$ experimental inputs.  

The RG running could be associated with the temporal evolution of the Universe
according to the identification $k \propto 1/H(t)$ where $H(t)$ is the
time-dependent Hubble-parameter, according to an idea promoted in
Ref.~\cite{alessia}. Thus, one could treat the field independent constant,
i.e., the $\Lambda$ term in Einstein's equations, as a running parameter which
varies over the temporal evolution of the Universe. If there is no other scalar
field in the theory, the cosmological constant must play an important role in
the mechanism of cosmic inflation, too.  So, its value must be fixed to be
large at a sub-Planckian scales. Thus, renormalization conditions must be
chosen very accurately in order to solve the corresponding
hierarchy problem. In other words, the running cosmological constant has to be
scaled down from high to low energies; this cannot be done straightforwardly in
the framework of conventional perturbative renormalization. So, again, a
nonperturbative treatment is needed. As an example, a nonperturbative approach
has been used in~\cite{Ru2020} to study the RG evolution in the presence of
compact extra dimensions. As a consequence, the studied parameters can be
substantially reduced to be comparable with observational values. So, if there
is a single scalar in the theory, then one would need to adjust the running of
the cosmological constant to accommodate both inflation and late-time dark
energy.  However, FRG generates infinitely many operators, so that together
with the cosmological constant the theory is equipped with an additional scalar
degree of freedom (the one of Starobinsky inflation) that can take care of the
inflationary era. This has been studied extensively in the literature, see for
instance \cite{AS_inflation_1,AS_inflation_2}.

Therefore, the use of the FRG method and asymptotic safety can solve several
important problems of cosmology and quantum gravity. Nevertheless, the FRG
approach has its own drawbacks; the evolution of the constant (i.e.,
field-independent) term could be problematic due to the construction of the
method. In particular, divergent terms could appear in the RG flow in its UV
limit. These could represent non-physical behaviour. It can be quite a subtle
problem to analyze the renormalization of the field-independent term, because
of the conditions imposed canonically on the regulator function in the FRG
equation (Wetterich equation)~\cite{eea_rg_1,eea_rg_2,Gi2006}. These
canonically imposed conditions imply that the solution of the FRG equation
reproduces the bare action in the UV only up to a field-independent, but
$k$-dependent term; the latter, in addition, could be infinite.  Hence, the
analysis becomes subtle. Yet, physical conditions, to be imposed on the RG
evolution of the constant term due to consistency considerations, lead to
subtracted RG equations whose solution fulfills all physical boundary
conditions in IR and UV limits.

Indeed, the idea to remove these UV divergent terms of the RG evolution of the
field-independent term in the framework of the FRG method is well known in
quantum-mechanical models, and in statistical mechanics. For example, as
suggested by quantum mechanical calculations [see Eq.~(37) of
Ref.~\cite{Gi2006} and additional discussions in
Refs.~\cite{Vacca_Zambelli,qanharmonic}], the application of a suitable
subtraction reduces the quadratic ($k^2$) divergence of the RG evolution in the
UV limit and produces correct results for the free energy in the IR limit.
However, it is very important to observe that the non-physical behaviour, i.e.,
the $k^2$ divergence of the UV limit is the consequence of the {\em absence} of
the Gaussian fixed point in the RG flow once the constant term is included. Of
course, one finds the Gaussian fixed point in the RG flow if the constant term
is explicitly set to zero and excluded from the 
model in its entirety. Once the constant term is included,
the problem of the $k^2$ divergence in the UV limit emerges,
and the Gaussian fixed point not only disappears;
it becomes unstable against tiny variations of the 
initial conditions of the RG flow. Conversely, if the
$\beta$-functions of a certain model allow the existence of the Gaussian fixed
point in the presence of the constant coupling, no further subtraction is
required.

It is possible to generalise the idea of the subtraction method for higher dimensions.
Indeed, a similar procedure has been proposed in Ref.~\cite{JCAP} for cosmological 
applications. In $d=4$ dimensions, in the absence of the Gaussian fixed point, $k^4$ and 
$k^2$ divergences appear in the RG flow of the field-independent but $k$-dependent term in 
its UV limit which may need special attention.

However, in the framework of the asymptotic-safety (AS) scenario, the situation is 
more complex. The $\beta$-functions of AS gravity  (without scalar 
degrees of freedom) predict the existence of Gaussian and non-Gaussian UV fixed 
points, see for example \cite{reuter,reuter_saueressig,bonanno_saueressig} 
and the RG flow diagram of \fig{fig0}.  At the non-Gaussian UV fixed point, the 
dimensionless couplings are required to attain a non-zero value. Thus, if we 
assume that the non-Gaussian UV fixed point is found in the asymptotic region, 
then at least some of the dimensionful counterparts of the couplings become 
divergent. For example, if we assume that the dimensionless cosmological constant 
$\lambda(k) = \Lambda(k) \, k^{-2}$ tends to a finite value at the non-Gaussian UV 
fixed point $\lambda(k = k_\star) = \lambda_\star$, then its dimensionful 
counterpart $\Lambda(k)$, for large $k$, can have at most a $k^2$ divergence. 
So, one could argue that a subtraction might be required which removes the $k^2$ 
divergence and leaves at most a logarithmic divergence, for large $k$.
However, AS gravity (without scalar fields) has the Gaussian UV fixed point, too. The 
$\beta$-functions derived from the FRG equation signal the Gaussian UV fixed point, 
so the field independent term vanishes if all the other couplings tend to zero in the UV 
limit. Thus, there is no need of any further subtraction method and the scaling of the 
potential as $k^4$ is not a problem in AS gravity since the $\beta$-functions predict 
the existence of the Gaussian UV fixed point, too.
However, if one adds scalar degrees of freedom as matter fields to the Einstein-Hilbert 
action which is the simplest truncation for AS gravity models, the resulting action, i.e., 
the simplest gravity-scalar system might have problem with the existence of the Gaussian
fixed point.  Of course, by the application of the subtraction method \cite{JCAP},
the Gaussian fixed point can be restored and the method, i.e., the inclusion of subtraction
terms does not affect the position of the non-Gaussian UV fixed point.

Therefore, the quantum field theory and the renormalization of gravity and also
the consequences of the RG running on cosmology are well defined, well
understood due to asymptotic safety (and the presence of the Gaussian fixed
point). However, one might ask what 
would happen if one chooses different approaches to the problem of
quantum gravity, for example M-theory or loop quantum gravity. In these cases,
one can use quantum field theory maybe together with supersymmetry and RG
considerations up to the Planck scale only. In addition, it is maybe required
to use an effective quantum field theory beyond the Standard Model.  As an
example one can mention the effective branon theory in (3+1) dimensions which
involves the absolute value of the branon field, leading to a
non-differentiable potential and wave function renormalization
\cite{non_diff_pot}. For discussions of branons, see, e.g.,
Refs.~\cite{Kugo,Bando,Dobado,Cembranos,branon_stabilization,Burnier,branon_dressing}.
This can re-open the discussion on the UV divergent nature of the RG flow of
the field-independent term, i.e., the cosmological constant which could play an
important role in the mechanism of cosmic inflation, too. For example, it is
known that supersymmetric extensions of the Standard Model deal with quadratic
divergences. It may resolve major hierarchy problems within the Standard Model,
by guaranteeing that quadratic divergences of all orders will cancel out in
perturbation theory, however, there is no experimental evidence that a
supersymmetric extension to the Standard Model is correct. Due to the negative
results from the LHC, it has been a matter of discussion whether the Minimal
Supersymmetric Standard Model is no longer able to fully resolve the hierarchy
problem. Thus, if not the asymptotic safety scenario is the correct approach to
quantum gravity, the problem of UV divergences of the field independent term
may require attention.

Indeed, in Ref.~\cite{MoSo2020}, the so-called Running Vacuum Model has been
studied; it assumes a scale-dependent vacuum, i.e., a running cosmological
constant. The authors of \cite{MoSo2020} proposed a suitable subtraction of the
so-called Minkowski contribution in the framework of the adiabatic
regularization method. It results in an RG scaling for the running vacuum
described by equation (6.9) of \cite{MoSo2020}, where the running RG scale is
denoted by the mass-like term $M$ while its UV value is given by $M_0$. The
result of their subtraction method is the absence of the $M^4$ term in the
scaling relation for the running vacuum. In the UV limit (when $M\to \infty$),
the $M^4$ is a source of exceedingly large contributions. Due to the
subtraction method of Ref.~\cite{MoSo2020}, the RG scale-dependence of the
running vacuum is characterised with $M^2$ and $\log(M^2)$ terms only. 
Our goal here is to show how the subtraction method can be further improved in
order to eliminate from the RG scaling of the running vacuum, not just the
$k^4$ (i.e., $M^4$ of Ref.~\cite{MoSo2020}) but also the $k^2$ (i.e., $M^2$ of
Ref.~\cite{MoSo2020}) terms which results in a purely logarithmic dependence on
the scale. 


\section{Field-Independent Term in Cosmology}
\label{sec3}

While, in ordinary particle physics, 
a constant, field-independent term of the potential 
carries no physical meaning, it has great importance
in the case of gravity. For example, 
in order to describe the observed accelerated expansion of the Universe 
at present \cite{accelerated_expansion_today_1,accelerated_expansion_today_2} 
a possible solution is the 
inclusion of a constant term into Einstein's equation which reads (in the 
absence of matter)
\begin{equation}
\label{einstein}
G_{\mu\nu} = R_{\mu\nu}-\hf R g_{\mu\nu} = - \Lambda g_{\mu\nu}
\end{equation}
where $R$ is the scalar curvature and $\Lambda$ is the cosmological 
constant. The latter is assumed to be related to dark energy 
\cite{dark_energy_1,dark_energy_2,dark_energy_3,dark_energy_4,dark_energy_5} 
and is expected to cause the accelerated expansion of the universe observed today. 
Indeed, by using the Friedmann--Lema\^{\i}tre--Robertson--Walker (FLRW,
see Ref.~\cite{FLRW_1,FLRW_2,FLRW_3,FLRW_4}) metric (in our units, the speed 
of light is $c \equiv 1$ and 
$\hbar \equiv 1$), $g_{\mu\nu} = \mr{diag}(-1,a^2,a^2,a^2)$ the scale factor $a(t)$ 
of the expanding homogeneous and isotropic Universe can be calculated which 
results in an exponentially fast expansion, $a(t) \sim \exp(\sqrt{\Lambda/3} \, t)$.

There is, however, a serious problem with this scenario, namely the discrepancy
between the theoretical prediction for the cosmological constant by quantum field 
theory (vacuum energy) and the energy density needed to explain the accelerated
expansion of the present Universe. In other words, the estimated value is 120 
orders of magnitude greater than the energy density of all the other matter.
Thus, the cosmological constant which correctly describes the acceleration rate
is very small compared to the Planck scale. This is the so-called cosmological 
constant problem.

One possible solution to this problem could be the existence of a hypothetical 
scalar field, referred to as quintessence 
\cite{quintessence_1,quintessence_2,quintessence_3,quintessence_4,quintessence_5,
quintessence_6,quintessence_7,quintessence_8,quintessence_9,quintessence_10},
which is minimally coupled to gravity. 
Compared to other scalar-field models such as 
k-essence \cite{k_essence_1,k_essence_2,k_essence_3}, 
quintessence is the simplest scenario where the slowly varying field along a 
potential results in a negative pressure and accelerated expansion. This 
mechanism is very similar to the particle physics model for cosmic inflation in 
the early Universe, but the difference is that non-relativistic matter cannot be 
ignored and the quintessence potential is much smaller than that of the inflaton 
potential. Let us first review the inflationary mechanism of the early Universe.

The key observation is that scalar fields can mimic the equation of state for
negative pressure. Thus they represent an excellent model for inflation,
\begin{equation}
\label{EH_and_matter}
S=\int d^4x \sqrt{-g} \left[\frac{m_p^2}{2} R +{\cal L}_\phi \right], \hskip 1cm
{\cal L}_\phi=-\hf g^{\mu\nu} \partial_\mu \phi \, \partial_\nu \phi -V(\phi) \,,
\end{equation}
where $\sqrt{-g}=\sqrt{-\det(g_{\mu\nu})}=a^3$ with the (reduced) Planck mass 
$m_p^2 = 1/(8\pi G)$, and $G$ is Newton's constant. 
The Einstein equation has to be written in the presence 
of matter fields,
\begin{equation}
G_{\mu\nu}= 8\pi G \, T_{\mu \nu},  \hskip 0.5cm  
T^\mu_\nu=\text{diag}(-\rho,p,p,p),
\end{equation}
where the stress-energy tensor of the scalar field has the following form,
\begin{equation}
\label{stessen}
T_{\mu\nu} = - \frac{2}{\sqrt{-g}} 
\frac{\delta(\sqrt{-g} {\cal L}_\phi)}{\delta g^{\mu\nu}}
= \partial_\mu \phi \, \partial_\nu \phi + g_{\mu\nu} \, {\cal L}_\phi.
\end{equation}
Since over the inflation the field can be considered to be homogeneous 
($\nabla \phi/a = 0$), the relation between the density and pressure reads as
\begin{equation}
\label{c1}
\omega=\frac{p}{\rho}=\frac{\hf \dot \phi^2-V}{\hf \dot \phi^2+V}
\hskip 0.5cm {\mr{if}}  \hskip 0.5cm 
\hf \dot \phi^2 \ll V \implies \omega=-1 \,,
\end{equation}
which results in exponential expansion similar to the case of the 
cosmological constant. Although the cosmological constant and the 
special equation of state ($\rho=-p$) both results in the same rate of 
expansion but the former cannot be used for inflation since it has to end. 
There is another condition for slow roll inflation which ensures a 
sufficiently prolonged inflation. These two slow-roll conditions in 
Planck units ($m_p \equiv 1$), have the following forms,
\begin{equation}
\label{conditions}
\epsilon \equiv \hat V'^{2}/(2 \hat V^{2}) \ll 1 \,,
\qquad
\eta \equiv \hat V''/\hat V  \ll 1 \,,
\end{equation}
which have to be fulfilled by a suitable potential for a prolonged 
exponential inflation with slow roll down. The e-fold number 
$N\equiv-\int_{\phi_i}^{\phi_f} d\phi \,\frac{V}{V'}$ should be in the range 
$50 < N < 60$ where $\phi_i$ and  $\phi_f$ are the initial and final 
configurations of the field, respectively. The power spectra of scalar 
(${\cal{P_S}}$) and tensor (${\cal{P_T}}$) fluctuations can be characterized 
by their scale dependence, i.e., ${\cal{P_S}} \sim k^{n_s -1}$, where $k$ is 
the comoving wave number. Then, slow-roll parameters are encoded in 
expressions for the scalar tilt $n_s-1 \approx 2 \eta -6 \epsilon$ and for 
the tensor-to-scalar ratio $r = {\cal{P_T}}/ {\cal{P_S}} \approx 16 \epsilon$, 
which can be directly compared to CMBR data \cite{planck_1,planck_2,planck_3}. 

The potential is determined by the slow-roll conditions \eq{conditions} 
up to an overall multiplicative factor, but this factor is fixed by the absolute 
normalisation. According to Eq.~(23) of Ref.~\cite{lyth_2} and Eq.~(218) 
of Ref.~\cite{baumann}, the normalisation condition is 
\begin{equation}
\label{norm_gut}
V(\phi_i) \equiv \frac{r}{0.01} (10^{16} \, \text{GeV})^4 \,.
\end{equation}
The tensor-to-scalar ratio $r$ is given by the slow-roll parameters 
which are fixed at the scale of inflation ($k_{\rminf}$),
according to remarks preceding Eq.~(218) of Ref.~\cite{baumann}. Thus, the
scale of inflation is given by the following relation
\begin{equation}
\label{Vcond}
V(\phi_i) \equiv k_{\rminf}^4 \,,
\qquad
k_{\rminf} = \left(\frac{r}{0.01}\right)^{\frac{1}{4}} 10^{16} \,\, {\rm GeV},
\end{equation}
which entirely fixes the inflationary potential including the constant term.
Therefore, the field-independent term is fixed at the scale of inflation, too.

Let us now come back to the idea of quintessence. The cosmological 
constant is static which means once it is fixed (e.g., to describe the 
rate of accelerating expansion of the Universe today) its value remains
constant over the history of the Universe. Thus, extrapolating back in 
time to the early Universe, it has a very small value compared to the
Planck mass. It would be more natural for the dark energy to start 
with an energy density similar to the density of matter and radiation 
in the early Universe. The concept of quintessence was introduced to 
overcome this problem which assumes a scalar field similar to inflaton
with negative pressure but with a very large wavelength. The equation 
of state of the quintessence is dynamic, time-evolving and given by
\eq{c1} which has been used for the inflaton case, too. Therefore, 
it does not matter whether one relies on the idea of quintessence
or on the cosmological constant. The field-independent term of the 
action is involved in the theoretical model,
and its value is determined. 

Thus, the Standard Model of Cosmology requires two periods of 
accelerated expansion: in the early Universe, when the Universe 
doubles in size in every $10^{-35}$s, and today, when the doubling 
time is 50 orders of magnitude greater. The field-independent term 
of the action has to be fine-tuned in each period of acceleration, thus
a reliable theory should take into account the change of this term 
over the time-evolution of the Universe. If the time-evolution of the
Universe and the momentum scale of renormalization is related to 
each other (inversely), then the required variation of the cosmological
constant in time can be produced by RG methods.

\begin{figure}[ht]
\begin{center}
\begin{minipage}{0.6\linewidth}
\begin{center}
\includegraphics[width=0.8\linewidth]{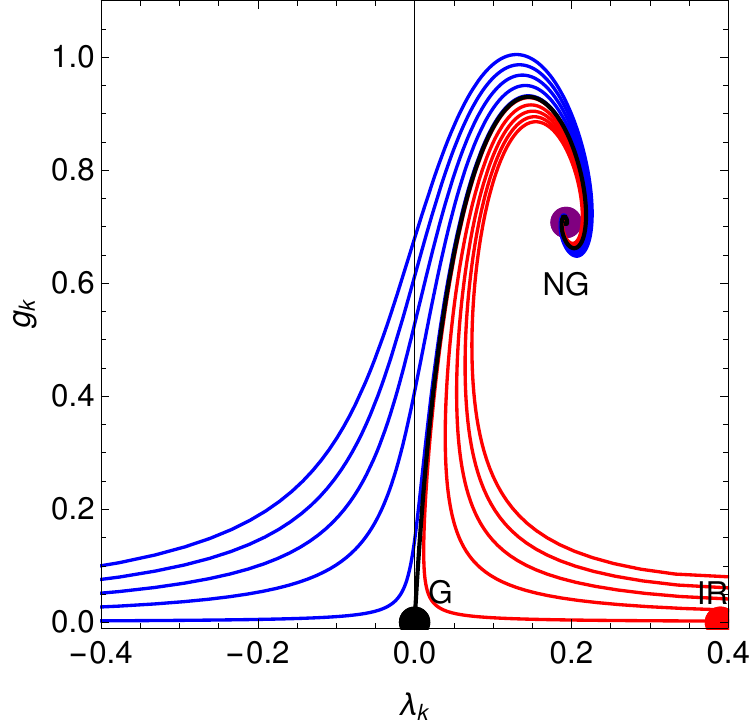}
\caption{
We present the RG flow diagram of QEG based on the Einstein-Hilbert 
trunction given in Eq.~\eq{EH}.
} 
\label{fig0}
\end{center}
\end{minipage}
\end{center}
\end{figure}

Indeed, the essential idea of the nonperturbative RG analysis is 
to describe the evolution of the self-interaction potential from the 
UV (from the Planck scale or from over the Planck scale) to the scale 
of inflation and towards to the low-energy (IR) limit, and of course, it is 
an important question how to describe the RG running of the constant term. 
The asymptotic safety scenario of quantum gravity is designed for that 
purpose, and its essence is the existence of a non-Gaussian UV fixed point. 
To summarise its cornerstones, one can start from the simplest realization 
of Quantum Einstein Gravity (QEG) which is the Einstein-Hilbert truncation
of the effective average action
\begin{equation}
\label{EH}
\Gamma_k = \frac{1}{16\pi G_k} \int d^4 x \, \sqrt{-g} \, 
(R - 2\Lambda_k) \,,
\end{equation}
where $g$ is the determinant of the metric tensor, $R$ is the Ricci scalar
and the scale-dependent parameters are the cosmological constant 
$\Lambda_k$ and the Newton coupling $G_k$. The field independent
term $V(\phi=0)$ of the scalar potential~\eq{EH_and_matter} is related 
to the cosmological constant, i.e., $V(\phi=0) = m^2_p \Lambda$. The 
scale-dependence of $\Lambda_k$ implies the scale-dependence of
$V_k(\phi=0)$, and it is analyzed in terms of dimensionless couplings, 
$\lambda_k \equiv \Lambda_k k^{-2}$, $g_k \equiv G_k k^2$ with the help of the 
$\beta$-functions, see for example \cite{bonanno_saueressig}
\begin{equation}
k \partial_k g_k = \beta_g \,, \hskip 1cm 
k \partial_k \lambda_k = \beta_\lambda
\end{equation}
which are calculated by the Litim regulator \cite{Litim2000}
\begin{equation}
\beta_g = (2 + \eta_N) g_k, \hskip 1cm \beta_\lambda = (\eta_N -2) \lambda_k 
+ \frac{g_k}{12\pi} \left[ \frac{30}{1-2\lambda_k} 
- 24 - \frac{5}{1-2\lambda_k} \eta_N \right] \,,
\end{equation}
where the anomalous dimension of Newton's constant 
$\eta_N = G_k^{-1} \, k\partial_k G_k$ 
is given by
\begin{equation}
\eta_N = \frac{g_k \, B_1}{1 - g_k \, B_2} \,,
\end{equation}
where
\begin{equation}
B_1 = \frac{1}{3\pi}\left[ \frac{5}{1-2\lambda_k} -
\frac{9}{(1-2\lambda_k)^2}  - 7 \right] , 
\hskip 1cm B_2 = 
- \frac{1}{12\pi}\left[ \frac{5}{1-2\lambda_k} -
\frac{6}{(1-2\lambda_k)^2} \right].
\end{equation}
The RG flow diagram based on the above $\beta$-functions is plotted on \fig{fig0}.
%
%

The $\beta$-functions contain the information on fixed points $g_\star$ and
$\lambda_\star$ of the RG flow where the beta functions vanish simultaneously.
They give rise to two fixed points: the Gaussian (G) UV fixed point situated at
$(g_\star, \lambda_\star) = (0,0)$ and the non-Gaussian (NG) UV fixed point
located at $(g_\star, \lambda_\star) = (0.707,0.193)$.  As we argued in the
introduction, the existence of the non-Gaussian UV fixed point can solve
important problems of quantum gravity. In order to find cosmological
applications, the running RG cutoff $k$ is identified with a typical length
scale of the system \cite{cosmo_qeg}. There are several types of cutoff
identifications \cite{cosmo_qeg}; among these,
one finds $k \sim t^{-1}$, where $t$ is the
cosmic time, or $k \sim H(t)$ where $H(t)$ is the Hubble parameter,
or $k \sim T$, where $T$ is the temperature of the cosmic plasma. The idea is to use RG
running to connect the physics of various energy scales. For example, one
should find the non-Gaussian fixed point above the Planck scale $k \gg m_p =  2.4 \times
10^{27}$ eV. By contrast, cosmic inflation takes place below the Planck scale $k =
k_{\rm{inf}} = 10^{22}$ eV, and the well-known value of Newton's constant is fixed
by laboratory experiments $G_k = G = 6.67 \times 10^{-57}$ eV$^{-1}$  at
low-energies $k = k_{\rm{lab}} = 10^{-5}$ eV. Finally, one should mention the
accelerated expansion of the Universe at present which requires $\Lambda_k =
\Lambda = 4 \times 10^{-66}$ eV$^2$ at the scale $k = k_{\rm{Hub}} = 10^{-33}$
eV. The nonperturbative RG (using various extension of the Einstein-Hilbert
truncation, see for example \cite{quadratic_gravity}) is capable to build up
connection between these scales and cover many orders of magnitude in change of
couplings, like the Newton and the cosmological constants. 

Our goal here is to consider the RG flow of the field independent
term in the presence of scalar fields, i.e. when the coupling constants of the
action \eq{EH_and_matter} are considered as running parameters similarly to
\eq{EH},
\begin{eqnarray}
\label{scalar_gravity}
\Gamma_k[\phi] =   \int d^4 x \, \sqrt{-g} \, \left[ \frac{1}{16\pi G_k} R
-\hf g^{\mu\nu} \partial_\mu \phi \, \partial_\nu \phi - V_k(\phi) \right] \,,
\end{eqnarray}
where the scalar potential is usually expanded in terms of the field. If this
expansion is terminated at the quartic order, it has the following form
\begin{eqnarray}
\label{scalar_gravity_pot}
V_k(\phi) = V_k(0) + \hf m_k^2 \phi^2 - 
\frac{1}{4!} \, \lambda_{4,k} \, \phi^4 \,, 
\hskip 1cm
V_k(0) \equiv \frac{2 \Lambda_k}{16\pi G_k} \,,
\end{eqnarray}
which is one of the simplest scenarios when a single real scalar field is
coupled to gravity.  For a detailed study of scalar fields coupled to
asymptotically safe quantum gravity see for example \cite{scalar_gravity} where
the existence of the non-Gaussian fixed point was shown for the simple case of
\eq{scalar_gravity_pot} with non-minimal coupling to gravity. The fixed points
of the RG flow for a scalar field in curved space with non-minimal coupling is
discussed in \cite{scalar_gravity_fixed_points}, too where the RG equation for
the scalar potential in the so-called Local Potential Approximation of the
Wetterich equation with the Litim cutoff reads as
\begin{eqnarray}
\label{potential_rg_equiation}
k\partial_k V_k(\phi) = \mu_d k^d \frac{k^2}{k^2 + \partial^2_\phi V_k(\phi)} \,,
\end{eqnarray}
with $\mu_d = 1/[(4\pi)^{d/2} \Gamma(d/2+1)]$.  The existence of the Gaussian
(G) and non-Gaussian (NG) fixed points and the RG flow of the full 
(dimensionless) potential is discussed, for example, in
Refs.~\cite{scalar_gravity} and \cite{scalar_gravity_fixed_points} but no
$\beta$ function is given for the field-independent term. In this work, we are
interested in the RG flow of the field-independent term, $V_k(0)$, with a
special attention on its UV limit.

Let us discuss the RG flow of of the cosmological constant in 
the absence of quantum gravity effects, but take into account the RG equation 
\eq{potential_rg_equiation} for the scalar potential \eq{scalar_gravity_pot}.
We use the relation $\tilde V_k(0) \equiv \frac{2 \lambda_k}{16\pi g_k}$ which
connects dimensionless couplings and gives
\begin{equation}
k\partial_k \lambda_k = 8\pi \left( g_k k \partial_k \tilde V_k(0) 
+ \tilde V_k(0) k\partial_k g_k\right)
\end{equation}
In the absence of quantum gravity effects, i.e., assuming a scale-independent 
dimensionful Newton's constant, $G_k = G$, the anomalous dimension vanishes 
because
$\eta_N = G^{-1} \, k\partial_k G =0$, and one finds $k \partial_k g_k = 2 g_k$ 
which results in a trivial RG scaling $g_k \sim k^{2}$. The RG flow equation for the 
dimensionless field-independent term obtained from \eq{potential_rg_equiation}
reads as
\begin{equation}
k \partial_k \tilde V_k(0) = \frac{1}{32 \pi^2} \left(\frac{1}{1+\tilde m^2_k} \right) 
- 4 \tilde V_k(0)
\end{equation}
which can be used to obtain the RG flow equation for the dimensionless cosmological
constant,
\begin{equation}
\label{flow_cosmo}
k \partial_k \lambda_k = \frac{1}{4 \pi} g_k \left(\frac{1}{1+\tilde m^2_k} \right) 
- 2\lambda_k. 
\end{equation}

In this approximation, $g_k$ has a trivial scaling, $g_k \sim k^2$, it tends to
infinity in the UV limit. The UV scaling of the dimensionless mass term is
$\tilde m^2_k \sim k^{-2}$, so, it tends to zero in the UV limit. Although the
UV Gaussian fixed point formally exists, it cannot be reached because the
corresponding $\beta$-function diverges in the UV limit, $g_k
\left(\frac{1}{1+\tilde m^2_k} \right) \to \infty$ if $k\to \infty$. Our goal
in this work is to show that one has to apply additional subtraction terms in
order to restore the Gaussian fixed point in \eq{flow_cosmo}.  

We will discuss that, even though the Gaussian fixed point is known to exist in
many quantum field theories, there may be questions regarding the proper
definition and retention of the Gaussian fixed point due to presence of the
divergent constant term.  A point emerging in our discussion is that with the
use of subtractions, one can restore the Gaussian fixed point of the pure
scalar theory in the RG flow based on the $\beta$-functions even if the
field-independent term is included. We will show that the inclusion of these
subtraction terms do not influence the non-Gaussian fixed point of the gravity-scalar
models, but could modify their IR behavior.

In summary, the constant term, i.e. the cosmological constant plays an
important role (i) in the physics above the Planck scale, (iii) in the
mechanism of inflation in the early Universe, (iii) in the accelerating
expansion at present. This justifies our main interest in the RG evolution of
the constant term in general, which we extensively discuss in the next section.

\section{Field-Independent Term in RG}
\label{sec2}

The modern formulation of nonperturbative RG usually referred as the
Wetterich RG equation \cite{eea_rg_1,eea_rg_2} has the following form for  the
one-component scalar field theory:
\begin{eqnarray}
\label{erg}
k \, \partial_k \Gamma_k[\phi] = \frac{1}{2} 
\int \frac{d^d p}{(2\pi)^d} \,
\frac{k \, \partial_k R_k(p)}{R_k(p) + \Gamma^{(2)}_k[\phi]}, 
\end{eqnarray}
where $k$ is the RG scale, $\Gamma_k[\phi]$ is the running effective action
with its Hessian $\Gamma^{(2)}_k[\phi]$, and $R_k(p)$ is the so-called regulator
function. It is illustrative to discuss its connection to the effective action,
which has the following form at the one-loop level:
\begin{eqnarray}
\label{effective_loop}
\Gamma_{\mathrm {eff}}[\phi] = 
S_\Lambda[\phi] + \frac{1}{2} \int \frac{d^d p}{(2\pi)^d} 
\ln\left[ S^{(2)}_\Lambda[\phi] \right]  + \ord{\hbar^2},
\end{eqnarray}
where $S_\Lambda$ is the classical (bare) action. A Pauli-Villars approach is
used to regularise the momentum integral which can be divergent at its upper
(UV) and lower (IR) bounds. This can be achieved by adding a momentum dependent
mass term $\hf \int R_k(p) \phi^2$ to the bare action, and introduce a
scale-dependent action
\begin{eqnarray}
\label{scale_effective}
\Gamma_k[\phi] \equiv  S_\Lambda[\phi] + \frac{1}{2} \int  \frac{d^d p}{(2\pi)^d} 
\ln\left[R_k(p) + S^{(2)}_\Lambda[\phi] \right] \,,
\end{eqnarray}
which recovers the effective action (at one-loop) in the IR limit if the regulator 
function $R_k(p)$ fulfils the requirements, $R_{k\to 0}( p) = 0$, $R_k(p\to 0) > 0$ 
[see Eqs.~(13)---(15) of Ref.~\cite{Gi2006}]. The latter condition is important
to avoid IR divergences. However, one canonically also imposes the condition
\begin{equation}
\label{regulator}
R_{k\to \Lambda}(p) = \infty
\end{equation}
(see Ref.~\cite{Gi2006}), and thus, in the UV limit, the scale-dependent action 
reproduces the classical (bare) action only up to a field-independent, constant 
term. If one can differentiate Eq.~\eq{scale_effective} with respect to the running 
scale $k$ (and multiplies both sides by $k$), then one finds
\begin{eqnarray}
k\partial_k \Gamma_k[\phi] = 
\frac{1}{2} \int  \frac{d^d p}{(2\pi)^d} 
\frac{k \partial_k R_k(p)}{R_k(p) + S^{(2)}_\Lambda[\phi]} \,,
\end{eqnarray}
which recovers the ``exact'' Wetterich RG equation \eq{erg} up to 
the replacement $S^{(2)}_\Lambda \to \Gamma^{(2)}_k$.

Let us come back to various limits of the scale-dependent action
\Eq{scale_effective}. It recovers the effective action in the limit $k \to 0$
and the bare action for $k \to \Lambda$, up to a field-independent but 
$k$-dependent term, which we will denote as $V_k(0)$ for reasons which will
become obvious immediately,
\begin{equation}
\label{GammaK2}
\Gamma_{k\to \Lambda}[\phi] = \Gamma_\Lambda[\phi] = 
S_\Lambda[\phi] + {\rm const.} =
S_\Lambda[\phi] + \int d^d x \, V_{k \to \Lambda}(0) \,.
\end{equation}
This clearly signals that the formulation of the RG evolution of the constant,
field-independent term $V_k(0)$ requires special care within the
nonperturbative approach implied by the Wetterich equation (see also 
Sec.~2.3 of Ref.~\cite{Gi2006}).  Moreover, if we implement the condition $R_{k\to
\Lambda}(p) = \infty$ on the regulator, then it turns out that in many cases,
the ``constant term'' $V_k(0) $ in Eq.~\eqref{GammaK2}, actually is given by a
divergent integral.

Therefore, the constant term $V_k(0) $ needs a special treatment in the
framework of the nonperturbative RG method.  In the following, we will
consider cases where $V_k(0)$ can naturally be identified with the
zeroth-order term (in $\phi$) obtained from the scale-dependent potential
$V_k(\phi)$.  One might argue that, for many purposes, the precise form of the
function $V_k(0)$ is physically irrelevant as it constitutes a
field-independent constant.  However, there are special cases where the RG
evolution of a constant (field-independent) part of the potential has physical
meaning. For example, if one aims at a determination of the free energy in a
flat background or of the cosmological constant in a general non-flat
background, then the problem of unambiguously determining $V_k(0)$ has to be
seriously considered.

The explicit form of the nonperturbative RG equation which is suitable for application
can be obtained by from~\eq{erg} using various approximations. Derivative
expansion is one of the widely used approximation and its leading order is the 
local potential approximation (LPA) \cite{Dupuis2021}. 
Within the LPA, the RG equation \eq{erg}, using the
so-called  Litim cutoff \cite{Litim2000}, reads as
\begin{equation}
\label{opt_dimful}
k\partial_k V_k =
\frac{2 \alpha_d}{d} \, \frac{k^{d+2}}{k^2+\partial^2_{\phi} V_k} \,,
\end{equation}
where $V_k$ is the dimensionful scaling potential in $d$ dimensions, $k$ is the 
running momentum, and $\alpha_d = \Omega_d/(2(2\pi)^d)$ is related to the 
$d$-dimensional solid angle $\Omega_d = 2 \pi^{d/2}/\Gamma(d/2)$. 
Eq.\,\eqref{opt_dimful} has been obtained projecting the exact functional RG equation\,
\cite{WeHo1973,Po1984,eea_rg_1,eea_rg_2,AlPo2001} [see Eq.\,\eqref{erg}] on 
a functional ansatz for the scalar field effective action, which contains only the bare 
kinetic term plus a scale dependent effective potential $V_{k}$.
Then, Eq.\,\eqref{opt_dimful} represents the non-perturbative $\beta$-function of 
the effective potential in absence of any renormalization for the non-local operators 
in the action\,\cite{BeTeWe2002,Po2003,Gi2006,Pa2007,De2007,Ro2012,Na2014}.

In the following, we will investigate the role of the RG evolution of the
field-independent terms in the framework of the nonperturbative RG approach.
We would like to use the RG treatment of the constant term developed for
low-dimensional quantum mechanical systems~\cite{Gi2006} and extend 
this method to higher dimensions.

\section{RG Evolution of the Constant Term}
\label{sec5}

We now come to the most important point to be 
discussed in the context of the current work.
Namely, from Eq.~\eqref{opt_dimful}, one could in 
principle derive an RG equation
for the constant term $V_k(\phi = 0)$.
Written for the dimensionful potential, one would
na\"{i}vely obtain the following RG equation
for $V_k(0)$ from Eq.~\eqref{opt_dimful} in $d=4$ dimensions:
\begin{equation}
\label{RGV0}
k\partial_k V_k(0) 
\mathop{=}^{\mbox{?}} 
\frac{k^4}{32\pi^2} 
\frac{k^2}{k^2+ \partial^2_\phi V_k(\phi) \vert_{\phi=0}} \,.
\end{equation}
It has already been mentioned above that,
within the nonperturbative RG equations used by us,
extra care is needed in the analysis of the field-independent terms.

The essence of the problem of the RG scaling of the 
constant term is the (possible) absence of the Gaussian
fixed point which, otherwise, is present, if the constant
term is not considered. In other words, the $\beta$-function
of the constant term should vanish if all couplings are set 
to zero. If the Gaussian fixed point is retained when the 
RG flow of the constant term is considered, which is the case 
in QEG, then one finds no problem with the RG scaling of the 
constant term. However, if the Gaussian fixed point is
missing once the field-independent coupling is included, 
then one finds problematic UV divergences which requires a
subtraction method. It was shown that such divergences 
occur for the quantum anharmonic oscillator and if one 
considers it in higher dimensions one has to generalise 
the subtraction method for $d=4$ dimensions.
Indeed, in the limit $k^2 \gg \partial_{\phi}^2 V_k(0)$,
the solution to Eq.~\eqref{RGV0} is 
\begin{equation}
\label{largek}
V_k(0)
\mathop{=}^{\mbox{?}} 
V_\Lambda(0)+\frac{k^4-\Lambda^4}{128\pi^2} \,,
\end{equation}
which would otherwise indicate a rampant quartic
divergence of $V_k(0)$ in the limit of large $\Lambda$
and lead to a considerable change in $V_k(0)$ between 
the Planck and the GUT scales, possibly requiring some 
fine-tuning of our model in the UV, i.e., at the Planck scale. 
In view of the quartic divergence of $V_k(0)$
in the UV, the fine-tuning problem would be 
of quite an extreme nature and conceivably 
render the model rather questionable.

In order to address the problem,
it is necessary to include a longer discussion.
The problem is exacerbated by the fact that
$V_k(0)$ could in principle be associated with a cosmological
constant term. As shown in the discussion
surrounding Eq.~\eqref{trafo1}---\eqref{scaling},
for an FLRW metric, 
one can bring the action into a form
resembling a flat-space theory,
but upon going back to the original FLRW coordinates,
one would realize that $V_k(0)$ indeed can take the role
of a cosmological constant [see also Eq.~\eqref{eq1_2}].

One should observe that the rampant 
$k^4$ behavior in Eq.~\eqref{largek} persists,
even if we set, e.g., all coupling terms
of the model to zero.
In principle, one might think that 
it would be difficult to argue that 
a quartic divergence could be obtained
for the RG running of the constant term $V_k(0)$ of 
a potential that completely vanishes in the IR.
The suspicion arises that the 
behavior implied by Eq.~\eqref{largek} 
cannot be physical and must be spurious.

\section{Divergences in $\maybebm{d=1}$ Oscillators}

We start this section by observing that 
questions related to the RG running of field-independent
constant terms belong to the more subtle questions
connected with nonperturbative RG equations.
In Sec.~2.3 of Ref.~\cite{Gi2006},
a $(0+1)$-dimensional model problem is studied which
illustrates the spurious nature of these terms,
namely, an anharmonic quantum mechanical oscillator.
For a potential of the form
\begin{equation}
V_k(x) =  E_{0k}  + \frac12 \, \omega_k^2 x^2 +
\frac{1}{24} \, \lambda_k^2 x^4 + \dots = \sum_{n=0}^{N_\text{cut}} \frac{ g_{2n,k}}{(2n)!}x^{2n},
\end{equation}
where $E_{0k} \equiv g_{0,k}$, $\omega^2_k \equiv g_{2,k}$ and $\lambda^2_k \equiv g_{4,k}$. 

One may investigate what happens if 
one drops the anharmonic term ($\lambda = 0$),
which implies that $\omega_k = \omega_{k=\Lambda} \equiv \omega$.
In this limit, the following RG equation is obtained
for the constant term 
$E_{0k}$ [see Eq.~(40) of Ref.~\cite{Gi2006}],
\begin{equation}
\label{NO}
\frac{d}{d k} E_{0k} \mathop{=}^{\mbox{?}}
\frac{1}{\pi} \, \frac{k^2}{k^2 + \omega^2} \,.
\end{equation}
When integrating this equation, one would obtain a
spurious linearly divergent ground-state energy 
of the anharmonic oscillator in the harmonic limit
($\lambda_k \to 0$, i.e., when the anharmonicity vanishes).

This behavior cannot be physical and cannot be trusted.
The answer is connected with the observation 
surrounding Eq.~\eqref{GammaK2}, which implies the necessity 
of subtracting a spurious field-independent constant
term if one would like to recover the bare action
(in the UV) from the nonperturbative RG equation.
In the case of the $(0+1)$-dimensional field theory,
the solution is given by subtracting the spurious 
term from the right-hand side RG equation~\eqref{NO},
and to solve instead
\begin{equation}
\label{YES}
\frac{d}{d k} E_{0k} \mathop{=}^{\mbox{!?}}
\frac{1}{\pi} \, 
\left( \frac{k^2}{k^2 + \omega^2} - 1 \right) \,.
\end{equation}
Integrating this equation, one obtains the 
correct ground-state energy $\frac12 \, \omega$
[see Eq.~(40) of Ref.~\cite{Gi2006}].

\begin{figure}[ht]
\begin{center}
\includegraphics[width=0.4\linewidth]{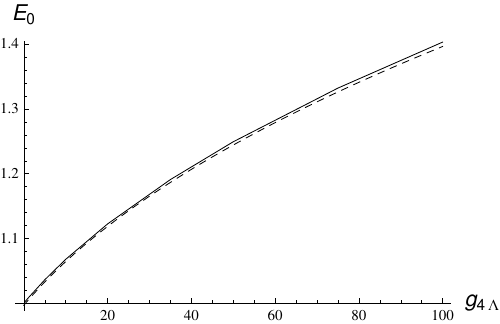}
\caption{
The dependence of the ground state energy of the quantum anharmonic oscillator on the 
$g_{4,\Lambda}$ initial condition, obtained for a single-well potential ($g_{2,\Lambda}=4$, 
$g_{0,\Lambda}=0$), using the truncation $N_{\text{cut}}=2$. The solid curve is from the 
solutions of the Schr\"odinger equations, while the dashed curve is computed by FRG.
} 
\label{fig1}
\end{center}
\end{figure}

\begin{figure}[ht]
\begin{center}
\includegraphics[width=0.4\linewidth]{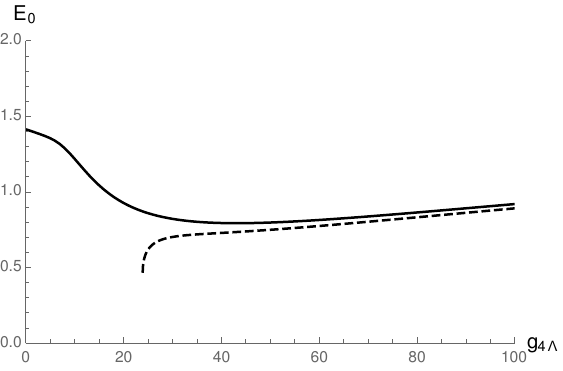}
\caption{
The dependence of the ground state energy of the quantum anharmonic oscillator on the 
$g_{4,\Lambda}$ initial condition, obtained for a double-well potential ($g_{2,\Lambda}=-4$, 
$g_{0,\Lambda}=\frac{3 g_{2,\Lambda}^2}{2g_{4,\Lambda}}$), using the truncation 
$N_{\text{cut}}=2$. The solid curve is from the solutions of the Schr\"odinger equations, 
while the dashed curve is computed by FRG.
} 
\label{fig2}
\end{center}
\end{figure}

Let us consider what happens if one keeps the anharmonic term ($\lambda \neq 0$). 
In order to study the dependence of the results on the truncation $N_\text{cut}$ we use 
the general form for the potential where the couplings are denoted by $g_{2n}$. One can 
derive the flow equations for the running couplings $g_{2n,k}$ by using the subtraction 
method explained above. For example by using the Litim regulator \cite{Litim2000}, and the truncation 
$N_\text{cut}=2$ the flow equations become
\begin{align}
k \partial_k g_{0,k}=&  \frac{k}{\pi} \left( \frac{k^2}{k^2+g_{2,k}}-1 \right),  \\
k \partial_k g_{2,k}=& -\frac{1}{\pi} \frac{k^3 g_{4,k}}{(k^2+g_{2,k})^2}, \\
k \partial_k g_{4,k}=& \frac{6}{\pi} \frac{k^3 g_{4,k}^2}{(k^2+g_{2,k})^3}.
\end{align}
By solving these differential equations we computed 
the ground state energy, $E_0=g_{0,k=0}$, with different initial conditions and got 
consistent results with the solutions of the Schr\"odinger equation as shown in \fig{fig1} 
and \fig{fig2}.

%
%
%

%
%

\begin{figure}[ht]
\begin{center}
\includegraphics[width=0.4\linewidth]{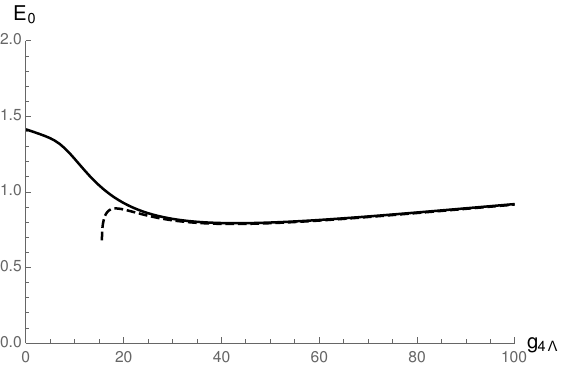}
\caption{
The dependence of the ground state energy of the quantum anharmonic oscillator on the 
$g_{4,\Lambda}$ initial condition, obtained for a double-well potential ($g_{2,\Lambda}=-4$, 
$g_{0,\Lambda}=\frac{3 g_{2,\Lambda}^2}{2g_{4,\Lambda}}$), using the truncation 
$N_{\text{cut}}=3$. The solid curve is from the solutions of the Schr\"odinger equations, 
while the dashed curve is computed by FRG.
} 
\label{fig3}
\end{center}
\end{figure}

The unphysical behaviour of the ground state energy for the double-well potential in low 
$g_{4,\Lambda}$ is just the result of our approximations. In fact better approximations improve 
these results, for example improving the truncation ($N_\text{cut}=3$) gives an improved curve, 
see \fig{fig3}.
%
%

Therefore, the use of the subtraction method for the set of RG flow equations recovers the 
correct results for the quantum anharmonic oscillator in $d=1$.

\section{Divergences in the Cosmological Model}

\subsection{General Discussion and Idea}

If the need for such subtractions arises in the 
context of $(0+1)$-dimensional field theories, 
then we can expect similar or even aggravated problems
in our $(3+1)$-dimensional case.
We should clarify that the subtraction leading from
Eq.~\eqref{NO} to Eq.~\eqref{YES} is not 
rigorously derived in Ref.~\cite{Gi2006},
but constitutes more than an {\em ad hoc} subtraction. 
Namely, as pointed out in the text preceding 
Eq.~(37) of Ref.~\cite{Gi2006},
the physical requirement is that the constant term
in the potential must remain zero under the RG
in the limit of vanishing parameters $\omega \to 0$,
and $\lambda \to 0$.
As pointed out in between Eqs.~(36) and~(37) of Ref.~\cite{Gi2006},
this requirement fixes the subtraction term.
The subtraction leading to Eq.~\eqref{YES} is 
justified by the fact that it reproduces
the known ground-state energy of the harmonic oscillator.

Based on the analogy with Ref.~\cite{Gi2006},
one might investigate if a
valid subtraction scheme for the field-independent
term in a nonperturbative RG could be 
obtained by subtracting from the naive
RG equation~\eqref{RGV0} the spurious 
asymptotic (UV) terms which cause the unphysical 
divergences. One should take into 
account that one single subtraction, as in
Eq.~\eqref{YES}, may not be enough.
In a different context, namely,
in the case of perturbatively renormalizable
theories such as quantum electrodynamics
(QED), one sometimes has to introduce
more than one subtraction term. E.g., the
regularization of the 
vacuum polarization integral, discussed
in Eq.~(7.3)~ff.~of Ref.~\cite{ItZu1980}
necessitates the introduction of more than one
``heavy fermion'' in order to eliminate 
a spurious quadratic divergence, for large loop
momenta.
In the case of the nonperturbative RG method
used here, one can show that, in general, 
the RG equation for $V_k(0)$ implies
a dependence with the functional form $k^d$ 
in the UV (when no subtractions are applied), 
consistent with the cases $d=1$ discussed
in Eq.~\eqref{NO} and the case $d=4$ discussed
in Eq.~\eqref{largek}.
This observation might suggest that 
more than one subtraction could be required
for higher dimensions.

Let us try to formulate possible,
but not unique, further requirements,
in terms of conjectures,
that could be imposed on the subtracted,
physical, RG evolution of the constant term.
These constitute mild generalizations
of the considerations reported in Ref.~\cite{Gi2006}.
\begin{itemize}
\item[\em (i)] The RG evolution of $V_k(0)$ must vanish
in the UV, in the limit of vanishing expansion coefficients
of the model.
\item[\em (ii)] The subtraction terms, 
originating from the UV, should have a polynomial 
functional form (in $k$), as they are obtained
from an asymptotic expansion of integrals 
obtained in the limit $k \to \Lambda$,
which typically gives rise to an asymptotic
expansion with polynomial terms.
\item[\em (iii)] The subtractions cannot possibly 
influence the IR behavior of the RG, 
as they originate purely from the UV behavior
of the regulator given in Eq.~\eqref{regulator}.
\end{itemize}
Condition {\em (iii)} implies that the subtractions
cannot induce IR divergences.
This means that one cannot subtract ``too many''
terms; otherwise one incurs infrared problems.

\subsection{Double Subtraction}

As suggested by the 
above general considerations, we observe that 
a single subtraction, based on the replacement 
\begin{equation}
\label{keep}
k\partial_k V_k(0)
\mathop=
\frac{k^4}{32\pi^2}
\frac{k^2}{k^2+\partial_{\phi}^2 V_k(0)} \to 
\frac{k^4}{32\pi^2}
\left[ \frac{k^2}{k^2+\partial_{\phi}^2 V_k(0)} - 1 \right] \; \to \;
- \frac{k^2}{32 \pi^2} \, \partial_{\phi}^2 V_k(0) \,,
\qquad k \to \infty \,,
\end{equation}
in Eq.~\eqref{RGV0}, leaves a quadratic term
(in $k$) on the right-hand side of Eq.~\eqref{RGV0}
in the UV limit. 
This subtraction term modifies the RG evolution of the cosmological 
constant. Thus, Eq.~\eq{flow_cosmo} is changed as
\begin{equation}
\label{flow_cosmo_1_subtraction}
k \partial_k \lambda_k = \frac{1}{4 \pi} g_k \left(\frac{1}{1+\tilde m^2_k} -1 \right)
- 2\lambda_k,
\end{equation}
from which one observes that the corresponding $\beta$-function does not
diverge in the UV limit but tends to a non-vanishing finite value,
i.e., the expression
$g_k \left(\frac{1}{1+\tilde m^2_k} -1 \right)$ approaches
a negative constant, where we keep in 
mind that $g_k \sim k^2$ and $\tilde m_k^2 \sim k^{-2}$ in the UV limit.

It is well known that comparable subtractions produce 
the correct result for the free energy of certain 
interacting quantum-statistical mechanics 
models~(see Chap.~8 of Ref.~\cite{Pet}).
Further remarks on related issues can also be found
in Ref.~\cite{KopietzLiebEtAl_1,KopietzLiebEtAl_2}.
One might argue that the negative value of $V_k(0)$
in the UV, implied if we assume Eq.~\eqref{keep}
to be valid in the UV, is akin to the negative mass square acquired 
by the Higgs particle upon considering tadpole diagrams
involving fermionic loops (Yukawa coupling), 
which also diverges quadratically
in $\Lambda$. Furthermore, one might argue
that this problem would be on the same level
as the so-called hierarchy problem of the Standard Model.
Conversely, if a subtraction is carried out in
the RG evolution of the field-independent term [see Eq.~\eqref{keep}], 
the hierarchy problem for 
the field-independent term in our model for the inflaton,
is reduced to the same level at which the Higgs particle mass of the Standard Model
itself suffers from a comparable RG running
and concomitant hierarchy problem.
(The lack of experimental evidence for supersymmetric 
particles, whose presence could potentially alleviate
the hierarchy problem of the Standard Model,
does not need to be stressed in the current context.)

The leading term on the right-hand side of 
Eq.~\eqref{keep} has a peculiar property:
It vanishes for vanishing couplings of the model 
but diverges for $\Lambda \to \infty$.
The physical condition of a vanishing RG evolution for the
field-independent term in the limit of vanishing
couplings is implemented 
in Eq.~\eqref{keep}, but the limit is
not approached uniformly in the sense that,
colloquially speaking, if $\Lambda$ goes to infinity
faster than the couplings go to zero, there is a remaining term. If, in
addition to conjectures {\em (i)}, {\em (ii)}, and {\em (iii)},
we also conjecture that the limit 
of vanishing parameters (expansion coefficients)
be approached smoothly, then the
second subtraction would be required.
Of course, this additional subtraction eventually has to be
justified on a calculation based on first principles.
It is permissible, though, to consider a
more favourable scenario where the use of an 
optimized regulator, which enables us to recover
the bare action from the solution of the RG equation in the UV,
will translate into a modified RG evolution for the 
constant term, which involves more than one subtraction
in the UV, for our four-dimensional case.
Let us therefore investigate the 
doubly-subtracted RG evolution
\begin{equation}
\label{RGV0sub}
k \partial_k V_k(0)
\mathop=
\frac{k^4}{32\pi^2}
\left( \frac{k^2}{k^2+\partial_{\phi}^2 V_k(0)} - 1 + 
\frac{\partial_{\phi}^2 V_k(0)}{k^2} \right) \,,
\qquad
k \to \infty \,,
\end{equation}
which is still infrared safe because of the 
$k^4$ prefactor. The subtraction of even more
asymptotic terms would contradict conjecture {\em (iii)}. 
In the UV limit, one then obtains
\begin{equation}
\label{RGV0sub_uv}
k\partial_k V_k(0) \sim
\frac{[ \partial_{\phi}^2 V_k(0) ]^2}{32\pi^2} \,,
\qquad
k \to \infty \,.
\end{equation}
{\em If} the double subtraction implied by Eq.~\eqref{RGV0sub} holds
for a modified RG evolution which avoids the unphysical divergences,
then this would lead to a cosmological constant that 
diverges in the UV, but only logarithmically.

Indeed, the RG flow equation \eq{flow_cosmo} of the dimensionless 
cosmological constant is modified as
\begin{equation}
\label{flow_cosmo_2_subtraction}
k \partial_k \lambda_k = \frac{1}{4 \pi} g_k 
\left(\frac{1}{1+\tilde m^2_k} -1 + \tilde m^2_k \right)
- 2\lambda_k,
\end{equation}
from which one observes that the $\beta$-function tends to zero in the UV
limit, since we have the asymptotic behavior that 
$g_k \left(\frac{1}{1+\tilde m^2_k} -1 + \tilde m^2_k\right) \to
0$ if $k\to \infty$, where we keep 
in mind that $g_k \sim k^2$ and $\tilde m_k^2 \sim
k^{-2}$ in the UV limit. 

In Ref.~\cite{MoSo2020} the authors show that no $M^4$ term is present (due to
the subtraction) and the RG scale-dependence is characterised by $M^2$ and
$\log(M^2)$ terms, where $M$ denotes the running RG scale.  Here we use a
double subtraction in \eq{RGV0sub_uv}, thus, there is no $k^4$ term in the UV
limit of the RG equation. If one neglects the scaling of the mass term $m_k \to
m$, the solution of the flow equation can be obtained and reads
\begin{equation}
\label{sol_double_rg}
V_k(0) \sim \frac{1}{32\pi^2} m^4 \log(k), \hskip 1.0cm k \to \infty.     
\end{equation}
Relating the RG scales $k \sim M$ in \eq{sol_double_rg} and by setting $\xi = 0$ 
in equation (6.9) of \cite{MoSo2020}, the two runnings can be compared, 
as is done in Sec.~\ref{compare}.

Let us
explore a {(possibly curious) analogy to the emergence of other logarithmic divergences
in quantum field theory, which seem to have a long history,
ever since Bethe, in his first calculation of the 
Lamb shift~\cite{Be1947}, obtained a logarithmically divergent
result (expressed in terms of a UV cutoff parameter),
in addition to the logarithmic sum over hydrogenic 
excited state which bears his name. 
The logarithmic 
divergence was classified as nonproblematic
in Ref.~\cite{Be1947}, as it was clear that a natural cutoff
(in the case of Ref.~\cite{Be1947}, the electron mass scale)
exists.  (In our case, of course, a natural UV cutoff is 
found at $k \sim \Lambda$, the Planck scale.)
The problem was later analyzed in greater detail
by French and Weisskopf~\cite{FrWe1949},
and Kroll and Lamb~\cite{KrLa1949},
as well as Feynman~\cite{Fe1949},
who clarified the matching of UV and IR divergences
(see p.~777 of Ref.~\cite{Fe1949}).
Without drawing any further analogies here,
we note that the original subtraction
introduced by Bethe in  Ref.~\cite{Be1947} was completely
{\em ad hoc} at the time; the full physical picture was clarified 
later. May it be permissible to 
mention that corresponding subtractions, to obtain physically
acceptable results in problems of advanced
classical electrodynamics, have recently been
discussed in Chap.~8 of Ref.~\cite{Je2017book}.

The subtractions introduced in Eqs.~\eqref{keep} 
or~\eqref{RGV0sub} are not unique,
and we do not have a rigorous derivation
at present beyond the considerations described above.
However, the physical requirements that should to be fulfilled
by the subtracted RG evolution of the constant term,
as formulated in the conjectures {\em (i)--(iii)} above,
are in agreement with the physical requirements
expressed between Eqs.~(36) and~(37) of Ref.~\cite{Gi2006}.
The conjecture and the subtraction terms enter
as an additional input into our calculations.
In general, it might be possible to avoid the unphysical
divergences of the constant, field-independent
terms via a suitable modification
of the condition~\eqref{regulator} imposed on the regulator function.

\subsection{Comparison with Other Results}
\label{compare}

In the previous subsection we proposed a new subtraction in order to get rid of
the spurious divergent terms $k^4$ and $k^2$ in the RG flow equation of the
constant (field-indepedent) term of the action, which is related to the
cosmological constant.  In the literature, the RG running of the cosmological
constant has been already suggested using subtraction ideas in order to solve
the cosmological constant problem.  Thus, let us compare them to our
double-subtracted RG equation \eq{RGV0sub_uv} and its solution.

In \cite{Ferrerio_Salas}, slightly generalized 
DeWitt-Schwinger adiabatic renormalization subtractions are proposed in
curved space to include an arbitrary renormalization mass scale $\mu$.
The running of the cosmological constant is obtained in equation (21) 
of Ref.~\cite{Ferrerio_Salas}. 
A similar running is given by Eq.~(14) of Ref.~\cite{Ferrerio_Salas_2},
which reads as
\begin{equation} 
\label{eq_Ferrerio_Salas}
\Lambda(\mu)=\Lambda_0 -\frac{1}{128 \pi^2} \left(-
(\mu^4-\mu_0^4) +2m^2 (\mu^2-\mu_0^2) 
-2m^4 \log \left( \frac{m^2+\mu^2}{m^2+\mu^4} \right) 
\right) \,,
\end{equation}
and it only differs from equation (21) of \cite{Ferrerio_Salas} in the prefactor 
of the second term of the right hand side by a multiplicative factor of 2.
The prefactor is, however, scheme dependent and it also depends 
on the physical content of the studied Lagrangian.
Thus, it is not surprising that one finds different 
values in the literature [see Eq.~(33) of \cite{Ferrerio_Salas_2}].
In Eq.~\eqref{eq_Ferrerio_Salas},
the running RG scale is denoted by $\mu$ with the UV value $\mu_0$.
Regarding the RG running, three distinct scale dependencies are present; 
a term that is proportional to $\mu^4$, one with $\mu^2$ and lastly a 
logarithmic term $\log{\mu^2}$. The RG running of the first two, 
especially of the $\mu^4$ term, can cause a rampant divergent behavior.

Now, let us take a look at the leading terms of Eq.~\eqref{RGV0} 
that do not yet contain subtractions. 
Insted of $\mu$ we denoted the RG running scale by $k$.
In the limit $k^2 \gg \partial_{\phi}^2 V_k(0)$,
the solution to Eq.~\eqref{RGV0} is Eq.~\eqref{largek},
\begin{equation}
V_k(0) \sim +\frac{1}{128\pi^2} k^4, \hskip 1.0cm k \to \infty,
\end{equation}
which indicates the same rampant quartic
divergence of $V_k(0)$. The prefactor $1/(128\pi^2)$ is also identical to the one in 
Eq.~\eqref{eq_Ferrerio_Salas}. 
If one uses a single subtraction \eq{keep} also proposed in \cite{Gi2006}, then the
leading term in the UV yields the solution
\begin{equation}
V_k(0) \sim -\frac{1}{64\pi^2} m^2 k^2, \hskip 1.0cm k \to \infty.
\end{equation}
This still contains a $k^2$ divergence in the UV limit, so, we propose a double subtraction, see equation \eq{RGV0sub_uv}, 
which has only a mild logarithmic divergence in the UV limit. For constant mass, its solution can be written as \eq{sol_double_rg}
\begin{equation}
V_k(0) \sim +\frac{1}{64\pi^2} m^4 \log(k^2), \hskip 1.0cm k \to \infty.   
\end{equation}
Again, the prefactors $1/(64\pi^2)$ also match exactly the prefactors of equation \eq{eq_Ferrerio_Salas}.

In Ref.~\cite{MoSo2020} the so-called Running Vacuum Model has been 
studied which assumes a scale-dependent vacuum, i.e, a running cosmological 
constant. The authors of \cite{MoSo2020} propose a suitable subtraction of the 
so-called Minkowski contribution in the framework of the adiabatic regularization 
method which results in an RG scaling for the running vacuum described by 
equation (6.9) of \cite{MoSo2020}
\begin{align}
\label{RGM}
\rho_{\rm vac}(M)=&\rho_{\rm vac}(M_0) + \frac{3}{16\pi^2} \left(\xi - \frac{1}{6} \right) H^2
\left[M^2-M_0^2-m^2 \ln \frac{M^2}{M_0^2} \right] \nonumber \\
&- \frac{9}{16\pi^2} \left(\xi - \frac{1}{6} \right)^2 
\left( \dot H^2-2H\ddot H-6H^2 \dot H \right) \ln \frac{M^2}{M_0^2} \,,
\end{align}
where the running RG scale is denoted by 
the mass-like term $M$ while its UV value is given by $M_0$. The result of the author's
subtraction method is the absence of the $M^4$ term in the scaling relation for 
the running vacuum. This has great importance; in the UV limit 
(when $M\to \infty$), the $M^4$ is a well-known source 
of exceedingly large contributions. 
The term $\xi$ stands for the non-minimal coupling to gravity.

One can relate the RG scale of 
Ref.~\cite{MoSo2020} 
[denoted as $M^2$ in Eq.~\eqref{RGM}] and the one used by us
(which we denote as $k$);
the result is the proportionality $k \propto M$.
Therefore, one can find very interesting
similarities between our approaches.  
In our case, we have minimal coupling, i.e., we do not include an extra $\xi R \phi^2$ 
coupling between gravity and the matter content, and thus $\xi = 0$.
Thus, for minimal
coupling, i.e., for $\xi = 0$, the numerical prefactors of the $M^2$ and
$\log(M^2)$ terms are $-1/(32\pi^2)$ and $1/(32\pi^2)$.
Despite the factor of 2 difference compared to our results,
we can say that the functional forms encountered in the RG
running in Ref.~\cite{MoSo2020} and here are compatible.
In Ref.~\cite{MoSo2020} the
authors show that no $M^4$ term is present (due to their subtraction of the Minkowskian contribution),
and the RG
scale-dependence is characterised by the $M^2$ and $\log(M^2)$ terms only.
Arguments for the absence of the quartic $M^4$ term are also presented in 
Refs.~\cite{Akhmedov,Ossola_Sirlin}, where similar flow equations are obtained.
If a double-subtraction method is used~\eq{sol_double_rg}, only
a logarithmic scale dependence remains.

\section{Conclusions}
\label{sec6}

In this work, we have discussed the role of the constant term in the
non-perturbative RG evolution.  In particular, we investigated whether the
rampant divergent terms $k^2$ and $k^4$ which naturally appear in the RG
equation for $d=4$ dimensions for the scalar inflaton field (in the absence of
the Gaussian fixed point) can be removed by a suitable subtraction method.
These divergent terms are the consequence of the construction of the functional
RG method and considered as unphysical.  They make the application of the
functional RG method on the proper treatment of the cosmological constant very
questionable if the Gaussian fixed point is absent. 

Renormalisation and RG scaling could be a possible tool to handle the
cosmological problem and they are under intense debate, see for example the
recent works, \cite{Ru2020,MoSo2020}.  In Ref.~\cite{MoSo2020} the Running
Vacuum Model has been studied which assumes a running cosmological constant
where the running vacuum described by equation (6.9) of \cite{MoSo2020} and the
running RG scale is denoted by the mass-like term $M$. Based on the subtraction
method of \cite{MoSo2020} it is possible to remove the divergent $M^4$ term and
RG scale-dependence of the running vacuum is characterised with $M^2$ and
$\log(M^2)$ terms only. 

Here we showed how the subtraction method can be improved further in order to
eliminate from the RG scaling of the running vacuum, not just the $k^4$ (i.e.,
$M^4$ of Ref.~\cite{MoSo2020}) but also the $k^2$ (i.e., $M^2$ of
Ref.~\cite{MoSo2020}) terms which results in a purely logarithmic dependence on
the scale, thus, the unphysical rampant behaviour of the naive approach is
handled.
As a future improvement on our subtraction method we
think it would be interesting to include the contributions of 
a theory that is non-minimally coupled to gravity ($\xi \neq 0$).  

\section*{Acknowledgments}

This work was supported by the \'UNKP-20-4-I New National Excellence Program of
the Ministry for Innovation and Technology from the source of the National
Research, Development and Innovation Fund and by the National Science Foundation
(Grant PHY--2110294).  Useful discussions with  Z.~Trocsanyi and G.~Somogyi are
gratefully acknowledged. This work was supported by the Deutsche 
Forschungsgemeinschaft (DFG, German Research Foundation) under Germany's Excellence 
Strategy EXC2181/1-390900948 (the Heidelberg STRUCTURES Excellence Cluster). 
The CNR/MTA Italy-Hungary 2019-2021 Joint Project "Strongly interacting systems in 
confined geometries" also is gratefully acknowledged.

\appendix

\section{Field--Independent Term and Cosmological Constant}
\label{app1}

The standard model of cosmology implies the requirement of an exponentially 
fast expansion~\cite{inflation,density-fluct_1,density-fluct_2,slow-roll_1,slow-roll_2} 
of the early Universe which 
is usually achieved by assuming a hypothetic inflaton field which slowly rolls 
down from a potential hill towards its minimum. Particle physics provides us with 
possible candidates for the inflaton field. A first guess involves a scalar field, 
$\phi$, as a candidate for the inflaton field, with a (Euclidean) action that 
includes the scalar curvature $R$ with minimal coupling to gravity,
\begin{equation}
\label{eq1_1}
S[\phi] = \int d^4 x \sqrt{-g} \left[ \hf 
g^{\mu\nu} \, \nabla_\mu \phi \, \nabla_\nu \phi + V(\phi) 
+ \frac{m^2_p}{2} R \right] \,,
\end{equation}
where the Planck mass $m^2_p \equiv 1/(8\pi G)$ has been used 
and $\sqrt{-g} = \sqrt{-\det g_{\mu\nu}} = a^3$ with 
$g_{\mu\nu} = \mr{diag}(-1,a^2,a^2,a^2)$.
Here, the scale factor $a = a(t)$ describes the cosmological scaling in the
Friedmann--Lema\^{\i}tre--Robertson--Walker (FLRW,
see Ref.~\cite{FLRW_1,FLRW_2,FLRW_3,FLRW_4}) metric (in our units, 
the speed of light is $c \equiv 1$ and $\hbar \equiv 1$). 

The FLRW metric is given by
\begin{equation}
\label{trafo1}
g_{\mu\nu} = \mr{diag}(-1,a(t)^2,a(t)^2,a(t)^2) \,,
\end{equation}
with the line element
\begin{equation}
\label{trafo2}
d s^2 = -dt^2 + a(t) \, d \vec r^{\,2} \,,
\end{equation}
where $\vec r = (x,y,z)$.
For scalar fields, the partial and covariant 
derivatives coincide, and hence, we can replace,
in Eq.~\eqref{eq1_1},
\begin{align}
\label{trafo3}
g^{\mu\nu} \, \nabla_\mu \phi \, \nabla_\nu \phi \to & \;
-\phi \nabla^\mu \nabla_\mu \phi 
= -\phi \left( -\frac{\partial^2}{\partial t^2} + 
\frac{1}{a(t)^2} \,  \frac{\partial^2}{\partial \vec r^2} \right) \, \phi \,.
\end{align}
One may stretch the spatial coordinates,
in a ``flattening'' transformation, according to
\begin{equation}
\label{scaling}
\vec r ' = a(t') \vec r \,,
\qquad
\phi(t, \vec r) = \phi'(t', \vec r') \,,
\qquad
t' = t \,.
\end{equation}
After this transformation, the action is brought into the 
form
\begin{equation}
\label{eq1_2}
S[\phi] = \int dt' \int d^3 r' \left[ -\hf \phi' \, \left( 
- \frac{\partial^2}{\partial t'^2} + \frac{\partial^2}{\partial \vec r'^2}
\right) \phi' + V(\phi') 
+ \frac{m^2_p}{2} R \right] \,,
\end{equation}
If we now re-identify 
\begin{equation}
t' \to i \, t \quad \mbox{(Wick rotation)} \,,
\qquad
\vec r' \to \vec r \,,
\qquad
\phi'(t', \vec r') = \phi(t, \vec r) \,,
\qquad
V(\phi') \to V(\phi) \,,
\end{equation}
and ignore the curvature term which is the cosmological constant term,
then we obtain
\begin{equation}
\label{Sapprox}
S[\phi] = \int d^4 x \, 
\left[ \hf (\partial_\mu \phi)^2 + V(\phi) \right] \,,
\end{equation}
with a fully local (Euclidean) action of our theory,
formulated in Euclidean four-dimensional space.
In comparison to Eqs.~\eqref{eq1_1},
Eq.~\eqref{Sapprox} has the same structure
as would otherwise be expected for flat space,
and makes the model amenable to a nonperturbative
RG analysis using established techniques~\cite{Litim2000}.

\end{document}